\documentclass[twocolumn,showpacs,preprintnumbers,amsmath,amssymb,prb,floatfix,superscriptaddress]{revtex4}

\usepackage{graphicx,tabularx}
\usepackage[version=3]{mhchem} 
\usepackage{amssymb}
\usepackage{dcolumn}
\usepackage{color}
\usepackage[mathcal]{euscript}
\usepackage{color}
\definecolor{gray0}{gray}{0.0}
\definecolor{gray64}{gray}{0.25}
\definecolor{gray128}{gray}{0.5}
\definecolor{gray192}{gray}{0.75}
\definecolor{gray255}{gray}{1.0}

\begin{document}
\title{Fingerprinting defects in diamond: Partitioning the vibrational spectrum}
\author{Danny E.P. Vanpoucke}
\affiliation{Institute for Materials Research (IMO), Hasselt University, 3590 Diepenbeek, Belgium}
\affiliation{IMOMEC, IMEC vzw, 3590 Diepenbeek, Belgium}
\email[]{Danny.Vanpoucke@UHasselt.be}

\date{\today}
\begin{abstract}
In this work, we present a computational scheme for isolating the vibrational spectrum of a defect in a solid. By quantifying the defect character of the atom-projected vibrational spectra, the contributing atoms are identified and the strength of their contribution determined. This method could be used to systematically improve phonon fragment calculations. More interestingly, using the atom-projected vibrational spectra of the defect atoms directly, it is possible to obtain a well-converged defect spectrum at lower computational cost, which also incorporates the host-lattice interactions. Using diamond as the host material, four test case defects, each presenting a distinctly different vibrational behaviour, are considered: a heavy substitutional dopant (Eu), two intrinsic defects (neutral vacancy and split interstitial), and the negatively charged N-vacancy center. The heavy dopant and split interstitial present localized modes at low and high frequencies, respectively, showing little overlap with the host spectrum. In contrast, the neutral vacancy and the N-vacancy center show a broad contribution to the upper spectral range of the host spectrum, making them challenging to extract. Independent of the vibrational behaviour, the main atoms contributing to the defect spectrum can be clearly identified. Recombination of their atom-projected spectra results in the isolated defect spectrum.
\end{abstract}

\maketitle
\section{Introduction}
Vibrational spectroscopy is an important tool for the structural investigation and characterization of solids.\cite{BonnotAM:SurfCoatTech1991,JacobCR:JChemPhys2009,ZivcovaZV:RSCAdv2016,LawD:Book2017,CorvaM:NatComm2018,WeckhuysenB:ChemEuroJ19,OzakiY:BullChemSocJpn2019} Quantum mechanical modeling of such experimental spectra starts from the calculated vibrational spectrum of the system, with the appropriate intensities for the individual spectral modes determined depending on the target experimental technique (\textit{e.g.}, Infra-Red, Raman,...).\cite{JacobCR:JChemPhys2009} Because of the inherent control over the atomic structure, such calculations can provide an invaluable source of information and understanding with regard to the structure of the system, and more specifically the impact of defects.\cite{SalustroS:JPhysChemA2018,VanpouckeDannyEP:2019c_DRM_EuDope, PetrettoG:CompMaterSci2018} However, the calculation of the vibrational spectrum of such defects faces a significant limitation: the calculation of an accurate and converged vibrational spectrum requires the use of large cells, which is computationally very demanding.\cite{PawelPT:JPhysChemLett2016,TeodoroTQ:JPhysChemLett2018} In the case of defects, the main interest goes to the modifications of the host spectrum due to the defect. It is therefore of interest not to spend computational resources on the reconstruction of the host spectrum, but to limit the calculations to the contributions to the spectrum due to the defect itself.\\
In contrast to solid state modeling, where the phonon spectra are generally only considered for small unit cell systems,\cite{TogoA:PhysRevB2015,PetrettoG:CompMaterSci2018} several approaches have been developed to deal with large systems within the context of (bio-)molecular structure investigations,\cite{SatoET:BiomedOptExpress18}. Examples are the selective calculation of specific normal modes,\cite{HerrmannC:NewJChem2007,TeodoroTQ:JPhysChemLett2018} the partial optimization of the molecular geometry,\cite{BourP:JChemPhys02} and the reduction of the Hessian by assuming rigid subsystems.\cite{HeadJD:IJQuantChem97,GhyselsA:JChemPhys2007,TerrettR:PCCP2017} Others have presented a fragment approach, in which a large system is decomposed in fragments of which high quality properties are calculated. The resulting fragment properties are then recombined again as approximation of the original system.\cite{BourP:JCompChem1997} Yamamoto \textit{et al.} \cite{YamamotoS:JChemTheorComp2012,YamamotoS:JPhysChemB2017} showed this method reproduces spectra of large molecules faithfully as long as suitable fragments were selected. They also note that the force field transfer accounted for nearly half of the observed error.\cite{YamamotoS:JChemTheorComp2012} Hanson-Heine \textit{et al.}\cite{HansonHeineMWDJChemTheoryComp2016} presented a local mode approach which can be used within the context of 2DIR spectroscopy of large systems, where it provides a platform for the parameterization of site frequencies and coupling maps with regard to the  geometry of different functional groups. Another proposed fragment strategy is centered on the construction of the Hessian matrix considering only the atoms in the region of interest.\cite{PascaleF:TheorChemAcc18} This approach efficiently succeeds in reproducing the spectra of interest, requiring only a small number of atoms to be considered. In this method, the selection of the atoms belonging to the fragment is rather \textit{ad hoc}. Furthermore, delocalized modes can not be treated with this approach. More specifically, within this approach the remainder of the system is kept frozen which may in some cases lead to unphysical behaviour if a normal mode is not tightly localized. A similar partial Hessian approach is the so-called Mobile Block Hessian approximation.\cite{GhyselsA:JChemPhys2007,TerrettR:PCCP2017} In this approach, the remainder of the system is also considered, but to reduce the computational cost, the atoms outside the fragment of interest are grouped in rigid blocks, which have no internal degrees of freedom, only $6$ external degrees of freedom. In contrast to these fragment approaches, Teodoro \textit{et al.} \cite{TeodoroTQ:JPhysChemLett2018} consider the full system using a computationally cheap approximate approach to obtain the initial full spectrum, after which only the normal-modes of experimental interest are selected and re-evaluated with a more accurate method.\\
In this work, we present a method to identify the atoms contributing to the vibrational spectrum of a defect. The overlap of the atom-projected spectrum with the reference host spectrum is presented as a suitable quantitative measure. We show that small supercells suffice to clearly identify the defect atoms and that the latter are limited in number.
We also show that the defect spectrum obtained through the combination of the atom projected spectra of the defect atoms is already well-converged when using a small supercell for the defect system. The resulting defect spectrum is continuous in nature, due to the incorporation of defect-host interactions. Four diamond based test cases are considered, for which the individual defect spectra are determined.\\
Within the context of the fragment approximations mentioned above, this method could resolve the \emph{ad hoc} nature of atom selection. Furthermore, within the context of the partial Hessian approximations, having a quantitative measure of the defect nature of an atom, would allow for more targeted selection of Hessian sub-blocks. In both cases, small supercells can be used to identify specific defect-atoms, and partial Hessian calculations on large supercells to obtain the spectrum of interest, reducing the computational cost of obtaining an accurate quantum mechanical vibrational spectrum in a periodic solid.\\

\section{Computational methods}
First-principles calculations are performed within the Density Functional Theory (DFT) framework using the VASP package.\cite{Kresse:prb99} The kinetic energy cutoff of the plane wave basis set is set to $600$ eV to obtain well converged forces, while the exchange correlation functional as proposed by Perdew, Burke and Ernzerhof (PBE) is used to describe the valence electron interactions.\cite{PBE_1996prl} The defects are imbedded in a $64$-atom conventional supercell, with the first Brillouin zone sampled by a $5\times 5\times 5$ Monkorst-Pack grid. Further details on the computational settings used are presented elsewhere.\cite{VanpouckeDannyEP:2017d_DRM_DFTUVac,VanpouckeDannyEP:2019c_DRM_EuDope}

\section{Harmonic phonon spectrum of solids}
In the following, we opted to use a very explicit notation for the dynamical matrix and its components over the more common and compact notation often found in the lattice dynamics literature.\cite{BornMHuangK1954:bookDynamicalTheoryCrystLatt} This was done with the aim of clarity, also for those less familiar with the field. Vectors are indicated in bold, and inner products are written as $\cdot$.
\subsection{Construction of the atom-projected phonon DOS}
\begin{figure}[tb!]%
\includegraphics*[width=\linewidth,keepaspectratio=true]{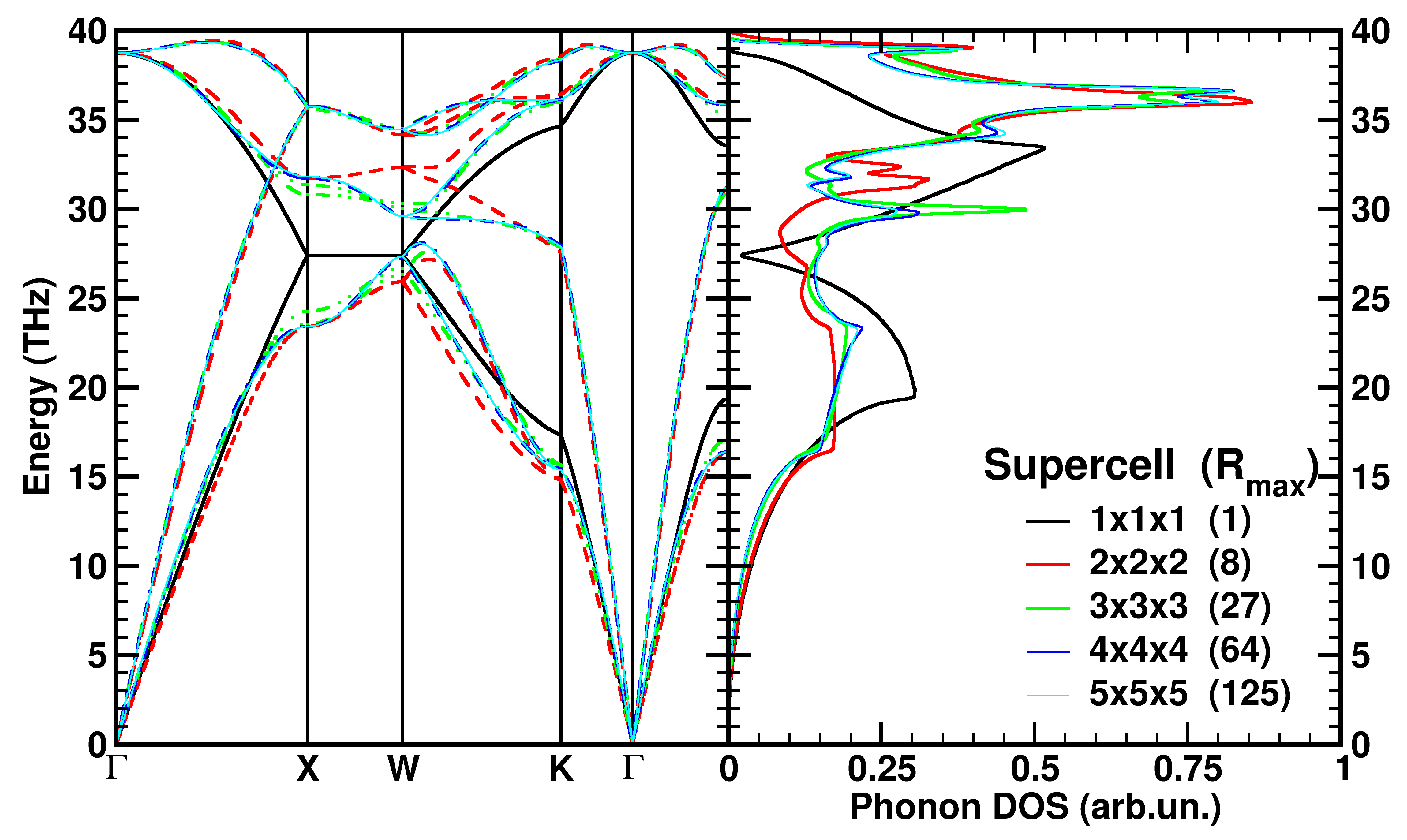}
\caption{%
  The vibrational band structure and resulting vibrational density of states (DOS) as obtained with Eq.~(\ref{eq_DynMat_qfull}) for pristine bulk diamond. Different color curves are used to show the convergence of the vibrational band structure and DOS with regard to the supercell size. The supercells are constructed starting from the primitive 2-atom unit cell. The supercells contain 2 ($1\time 1\times 1$), 16 ($2\time 2\times 2$), 54 ($3\time 3\times 3$), 128 ($4\time 4\times 4$) and,  250 ($5\time 5\times 5$) atoms.
\label{fig_SpectrConv}}
\end{figure}

Most modern quantum mechanical and quantum chemistry packages provide access to the vibrational spectrum of a system at the center of the first Brillouin zone. This vibrational spectrum at the $\Gamma$-point, can be obtained by the diagonalisation of the mass-weighted Hessian matrix, also called dynamical matrix:
\begin{equation}\label{eq_DynMat_Gam}
  D_{mol}(\Gamma)=\begin{bmatrix}
            \frac{\varphi(N_1,N_1)}{\sqrt{m_1m_1}} & \frac{\varphi(N_1,N_2)}{\sqrt{m_1m_2}} & \cdots & \frac{\varphi(N_1,N_n)}{\sqrt{m_1m_n}} \\
            \frac{\varphi(N_2,N_1)}{\sqrt{m_2m_1}} & \frac{\varphi(N_2,N_2)}{\sqrt{m_2m_2}} & \cdots & \frac{\varphi(N_2,N_n)}{\sqrt{m_2m_n}} \\
            \vdots & \vdots & \ddots & \vdots \\
            \frac{\varphi(N_n,N_1)}{\sqrt{m_nm_1}} & \frac{\varphi(N_n,N_2)}{\sqrt{m_nm_2}} & \cdots & \frac{\varphi(N_n,N_n)}{\sqrt{m_nm_n}}
         \end{bmatrix},
\end{equation}
with $N_a$ indexing the atoms of the system, $m_a$ the atomic mass, and the $3\times 3$ matrices $\varphi(N_a,N_b) = [\varphi_{i,j}(N_a,N_b)]$, $i,j=x,y,z$. The individual matrix elements $\varphi_{i,j}(N_a,N_b)$ can be determined either as the second derivative of the total energy with regard to the displacements $x_i(N_a)$ and $x_j(N_b)$ of atom $N_a$ and $N_b$, respectively, or as the derivative of the forces acting on atom $N_a$ by the displacement of atom $N_b$:\cite{AcklandGJ:JPhysCondMat1997}
\begin{equation} \label{eq_Hessian}
\varphi_{i,j}(N_a,N_b) = \frac{\partial^2 E}{\partial x_i(N_a)\partial x_j(N_b) } = -\frac{\partial F_i(N_a)}{\partial x_j(N_b)}.
\end{equation}
For molecular systems, or clusters, a dynamical matrix as presented in Eq.~(\ref{eq_DynMat_Gam}) would provide a complete picture. However, in the case of a periodic solid there are two complications that need to be dealt with: (1) the infinite nature of a theoretical crystal and (2) the finite size of the first Brillouin zone.\\
For a periodic crystal, all relevant physics is contained in a single unit cell, reducing the number of atoms to consider from infinity to a (small) finite number. On the other hand, to obtain the phonon density of states of a periodic solid, one needs to integrate the spectrum over the full Brillouin zone (similar as for the calculation of the electronic density of states). As such, one infinity is replaced by another, albeit a more manageable one. The vibrational spectrum at each point $\mathbf{q}$ of the Brillouin zone (BZ) is determined through the diagonalisation of the dynamical matrix:\cite{BornMHuangK1954:bookDynamicalTheoryCrystLatt}
\begin{equation}\label{eq_DynMat_q}
  D_{BZ}(\mathbf{q})=\begin{bmatrix}
            \frac{\varphi(N_1,N_1)}{\sqrt{m_1m_1}}e^{i\mathbf{q}\cdot(\mathbf{r}_{N_1}-\mathbf{r}_{N_1})} & \frac{\varphi(N_1,N_2)}{\sqrt{m_1m_2}}e^{i\mathbf{q}\cdot(\mathbf{r}_{N_1}-\mathbf{r}_{N_2})}  & \cdots \\
            \frac{\varphi(N_2,N_1)}{\sqrt{m_2m_1}}e^{i\mathbf{q}\cdot(\mathbf{r}_{N_2}-\mathbf{r}_{N_1})}  & \frac{\varphi(N_2,N_2)}{\sqrt{m_2m_2}}e^{i\mathbf{q}\cdot(\mathbf{r}_{N_2}-\mathbf{r}_{N_2})}  & \cdots \\
            \vdots & \vdots & \ddots
         \end{bmatrix},
\end{equation}
with $\mathbf{r}_{N_a}-\mathbf{r}_{N_b}$ the real space vector from atom $b$ to atom $a$.\\
Furthermore, because interatomic interactions are infinitely ranged, the dynamical matrix needs to incorporate interactions with other unit cells as well. Indexing the unit cells, with $R=1$ being the reference unit cell (UC), the general form of the dynamical matrix can be written as:
\begin{equation}\label{eq_DynMat_qfull}
  D_{BZ,UC}(\mathbf{q})=\sum_{R=1}^{\infty}{ \begin{bmatrix}
            \frac{\varphi(N_1^1,N_1^R)}{\sqrt{m_1m_1}}e^{i\mathbf{q}\cdot(\mathbf{r}_{N_1^1}-\mathbf{r}_{N_1^R})} & \frac{\varphi(N_1^1,N_2^R)}{\sqrt{m_1m_2}}e^{i\mathbf{q}\cdot(\mathbf{r}_{N_1^1}-\mathbf{r}_{N_2^R})}  & \cdots \\
            \frac{\varphi(N_2^1,N_1^R)}{\sqrt{m_2m_1}}e^{i\mathbf{q}\cdot(\mathbf{r}_{N_2^1}-\mathbf{r}_{N_1^R})}  & \frac{\varphi(N_2^1,N_2^R)}{\sqrt{m_2m_2}}e^{i\mathbf{q}\cdot(\mathbf{r}_{N_2^1}-\mathbf{r}_{N_2^R})}  & \cdots \\
            \vdots & \vdots & \ddots
         \end{bmatrix} }.
\end{equation}
For practical purposes, $R$ can be truncated to a finite number of unit cells, $R_{max}$, as the contributions to the dynamical matrix of unit cells farther away becomes vanishingly small.\cite{BornMHuangK1954:bookDynamicalTheoryCrystLatt} The convergence of the vibrational band structure and density of states (DOS), as function of $R_{max}$, is shown in Figure~\ref{fig_SpectrConv}. Note that diamond has a rather small primitive unit cell. For large unit cell systems, such as for example metal-organic frameworks,\cite{VanpouckeDannyEP:2015b_JPhysChemC} a converged spectrum may be obtained already at the unit cell level (\textit{i.e.}, $R_{max}=1$).\\
Note that for supercell calculations, which are generally used to obtain vibrational spectra from quantum mechanical calculations,\cite{QainX:PhysRevB2018} the supercell dynamical matrix, Eq.~(\ref{eq_DynMat_q}), and the unit cell dynamical matrix, Eq.~(\ref{eq_DynMat_qfull}), are related through symmetry. Both give rise to the same phonon DOS, however, as matrix diagonalization scales approximately as $O(n^3)$, Eq.~(\ref{eq_DynMat_qfull}) is much more efficient for larger supercells. This is of interest when using a dens sampling of the BZ.\\

The dynamical matrix is diagonalised by solving the following eigenvalue problem:
\begin{equation}\label{Eq_eigval}
  D(\mathbf{q})\cdot \mathbf{v}(\mathbf{q},j)=\omega^2(\mathbf{q},j) \mathbf{v}(\mathbf{q},j),
\end{equation}
with $\omega^2(\mathbf{q},j)$ the $j^{\mathrm{th}}$ eigenvalue at wave vector $\mathbf{q}$ and $\mathbf{v}(\mathbf{q},j)$ the corresponding eigenvector. This eigenvector $\mathbf{v}(\mathbf{q},j)$ represents the mass-weighted displacement vectors associated with phonon-mode $j$. From this it is possible to construct a weighing for each atom:
\begin{equation}\label{Eq_AtWeight}
  w_a(\mathbf{q},j) = \frac{\sum_{i=x,y,z}{|\mathbf{v}_{a_i}(\mathbf{q},j)|^2}}{\|\mathbf{v}(\mathbf{q},j)\|^2},
\end{equation}
allowing for a partitioning of the phonon DOS. The weighing factors normalize to one as $\sum_a{w_a(\mathbf{q},j)=1}$.\cite{note_EigVectNorm} The atom-projected phonon spectrum for atom $a$ at a frequency $\nu$ is then calculated as:
\begin{equation}\label{Eq_atProjDos}
  \omega_a(\nu) = \sum_{j=1}^{3N}{\frac{1}{V_{BZ}}\int_{BZ}{\omega(\mathbf{q},j)w_a(\mathbf{q},j)\delta(\nu,\omega(\mathbf{q},j)) d\mathbf{q}}},
\end{equation}
with $V_{BZ}$ the volume of the first Brillouin zone and $\delta$ the Dirac delta function.\\

\subsection{Differences of spectra}
When trying to extract the part of the vibrational spectrum due to a defect, one may be tempted to take the difference of this full spectrum and a reference spectrum (\textit{i.e.}, the spectrum of the host material). The result will contain clear defect features---such as new peaks outside the host spectrum, and new intense features within the range of the host spectrum. This can provide a reasonable qualitative picture, even though significant noise as well as negative intensities are to be expected.\\
Correct normalisation with regard to the host spectrum is complicated by the possible difference in number of atoms (\textit{e.g.}, due to an interstitial or vacancy), but also by the question of which atoms belong to the defect (\textit{e.g.}, only the substitutional dopant, or also nearest neighbours?).\\
To move beyond the qualitative identification of defect related vibrational states and properties it is necessary to obtain a well-normalised spectrum (\emph{i.e.}, the integrated DOS corresponds to the actual number of states involved in the defect spectrum) as well as an associated listing of defect-contributing atoms. The resulting defect-host partitioning should be independent of the defect system, and provide a quantitative measure for identifying atoms belonging to the defect.

\subsection{Isolating the phonon spectrum of a defect}
\begin{figure}[tb!]%
\includegraphics*[width=\linewidth,keepaspectratio=true]{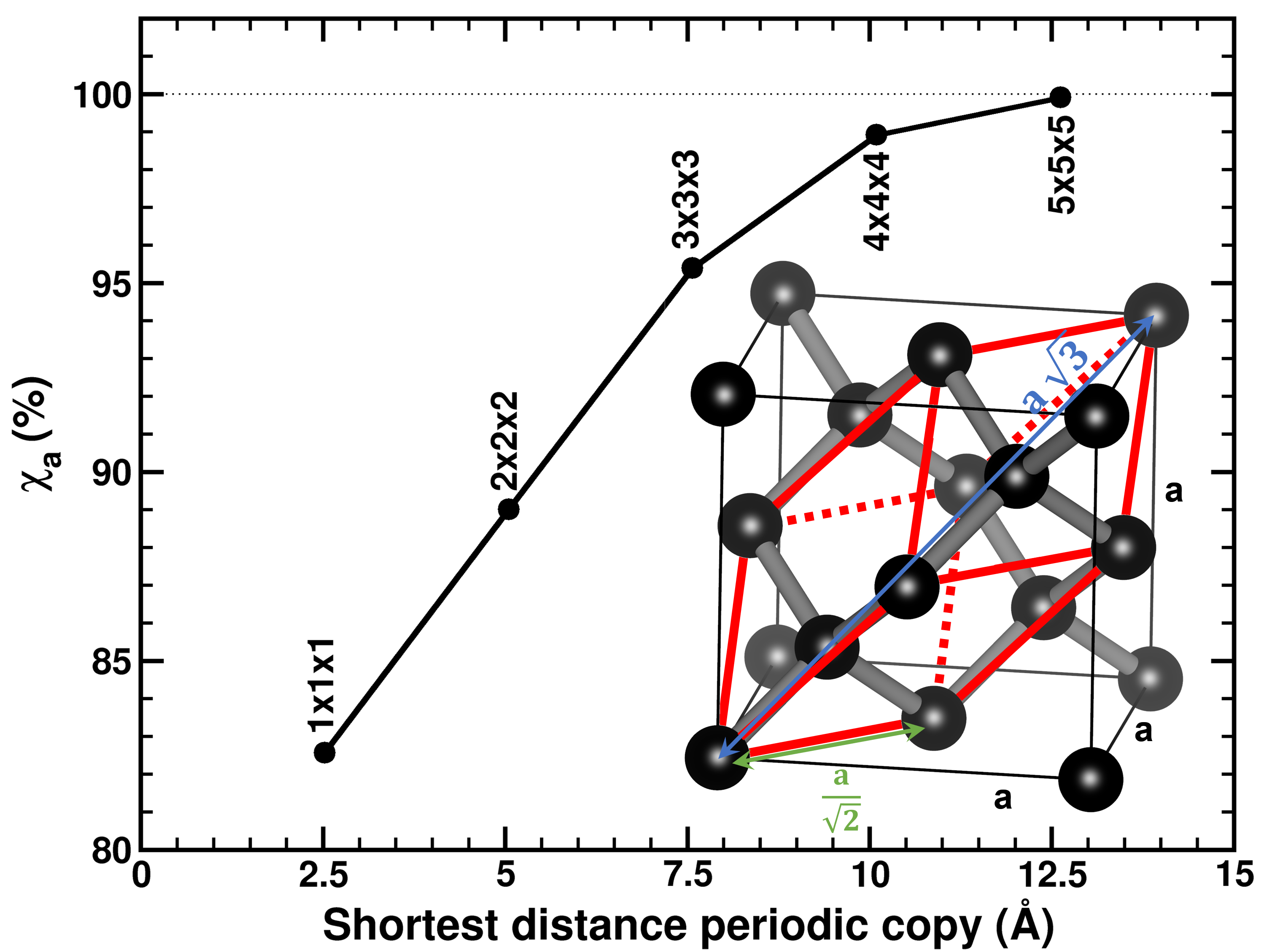}
\caption{%
  Convergence of the diamond vibrational spectrum as function of the supercell size, using $\chi_{C}$ as quality measure. The vibrational DOS of the $5\times 5\times 5$ supercell is used as reference.\label{fig_RMSDoverlap} }
\end{figure}

\begin{figure}[tb!]%
\includegraphics*[width=\linewidth,keepaspectratio=true]{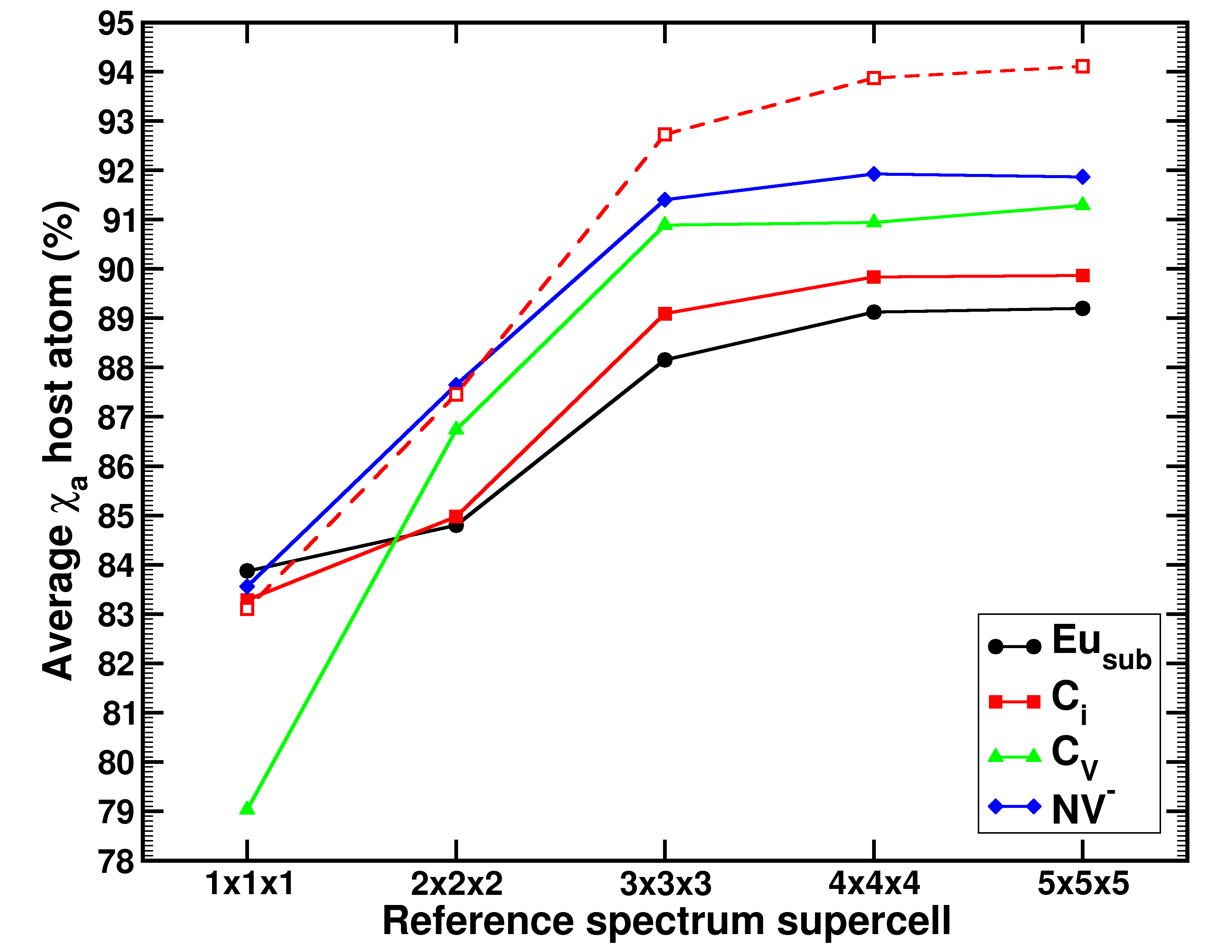}
\caption{%
  Convergence of $\chi_a$ of C host atoms in the different defect systems as function of supercell used in creating the reference host spectrum.
  The overlap is calculated as the average of the overlap $\chi_a$ for all atoms $3.0$--$5.0$ \AA\ removed from the center of the defect in the $64$ atom conventional supercells. The empty red squares show the values obtained for the C$_i$ defect in a $216$ atom conventional supercell, for comparison.\label{fig_overlapConvergence} }
\end{figure}

In this work, we address the problem of isolating the phonon spectrum of a defect, starting from the uncertainty of which atoms contribute to the defect spectrum. Although the approach can be extended easily to host materials with multiple inequivalent atomic sites and multiple atomic species, we present the methodology from the perspective of a host material containing only one atomic species and one inequivalent atomic position: diamond. Since all atoms are perfectly equivalent in this system, their contribution to the phonon spectrum is exactly the same (\textit{i.e.}, $\frac{1}{n}$ times the total phonon spectrum of a unit cell containing $n$ atoms). Therefore, the impact of a defect can be identified clearly as the deviation from this reference spectrum.\\
A straightforward method to quantify this difference is by means of the Root-Mean-Square-Deviation (RMSD) of the two (normalized) spectra:
\begin{equation}\label{Eq_RMSD}
  RMSD(a) = \sqrt{\frac{\int_{\nu=0}^{\nu_{max}}{(\tilde{\omega}(\nu) - \tilde{\omega}_a(\nu))^2 d\nu}}{\nu_{max}}},
\end{equation}
with $\nu_{max}$ the highest frequency of the spectra $\tilde{\omega}$ and $\tilde{\omega}_a$, the normalised host and atom spectrum. In this case, a host atom has theoretically an RMSD of zero. A defect atom, on the other hand, has a positive non-zero RMSD. However, as the upper bound of this function strongly depends on the shape of the spectra $\tilde{\omega}$ and $\tilde{\omega}_a$, the information gained is too limited for our purpose.\\
Alternately, the overlap of the (normalized) phonon spectrum obtained for the host system and atom $a$ of the defect system presents a bound function with an upper value of $1$ (or 100\%):
\begin{equation}\label{Eq_overlap}
  \chi_a = \Bigg(\int_{\nu}{\min{(\tilde{\omega}(\nu),\tilde{\omega}_a(\nu))}d\nu}\Bigg)\times 100\%.
\end{equation}
In this case, a \emph{host atom} shows $100$\% overlap while a \emph{defect atom} shows a lower value. A value of $0$\% could in theory be obtained for a defect atom which gives rise only to vibrational contributions outside the host spectrum. The substitutional Eu atom in diamond, which we will discuss later, approaches this theoretical limit with $\chi_{Eu} = 10$\%.\\
We noted earlier that the vibrational spectrum and DOS for systems with a small unit cell (such as prospective host systems) may require rather large supercells to present a converged picture (\textit{cf.}, Figure~\ref{fig_SpectrConv}). Defects, in contrast, are modelled using large supercells to approximate the experimentally relevant ``low'' defect concentrations. As a result, longer ranged vibrational interactions are by default incorporated for such systems, leading to a more converged phonon spectrum than is the case for a small host system unit cell (\textit{cf.}, black curve in Figure~\ref{fig_SpectrConv}). It is therefore essential to obtain a sufficiently converged host \emph{reference} spectrum,$\tilde{\omega}(\nu)$ , (\textit{cf.}, the $5\times 5\times 5$ spectrum in Figure~\ref{fig_SpectrConv}) to avoid artificial overlap mismatch when calculating $\chi_a$.\\
This mismatch can be quantified by calculating the convergence of $\chi_a$ of the host spectrum itself. In the case of diamond, this is shown in Figure~\ref{fig_RMSDoverlap}. \textit{E.g.}, the $\chi_a$ of a 2-atom diamond system gives a mismatch of about $20$\%. As such, a ``host atom'' in a defective system, when using the 2-atom reference data, would have a $\chi_a$ of about $80$\%, instead of the theoretical maximum of $100$\%. This is seen in Figure~\ref{fig_overlapConvergence}, showing $\chi_a$ of the host C atoms in the defect systems to be in the range of $79$-$84$\% for the $1\times 1\times 1$ reference spectrum. Fortunately, the computational cost of obtaining well converged reference host spectra is not extreme when taking advantage of (translational) symmetry. We therefore assume in the following that the reference spectra are sufficiently converged.\\
For the ``host atoms'' in defect systems, long ranged interactions will impact their expected $\chi_a$ as well. As can be seen in Figure~\ref{fig_overlapConvergence}, this leads to a leveling of the host atom $\chi_a$ as function of the reference spectra used.

For a defect modeled with a $64$ atom conventional supercell (which is relatively small for a defect cell), the $\chi_a$ of the host atoms ranges between $89$ and $92$\%, which is consistent with the convergence of the host atom overlap in the primitive $2\times 2\times 2$ cell (\textit{cf.}, Figure~\ref{fig_RMSDoverlap}).
For larger defect cells, the overlap of the host atoms increases, as can be seen for the example of the C$_i$ defect (\textit{cf.}, Figure~\ref{fig_overlapConvergence}). Placing the defect in a conventional $3\times 3\times 3$ supercell, $\chi_a$ increases to about $95$\%, which in turn, is consistent with the convergence of the reference host atom in the primitive $3\times 3\times 3$ cell (\textit{cf.}, Figure~\ref{fig_RMSDoverlap}).  More interestingly, this increase is not due to a gradual increase in $\chi_a$ for atoms ever farther away from the defect, but rather due to a general upward shift of the overlap of the \textit{non-defect} atoms, as can be seen in Figure~\ref{fig_overlapVSdist}.

In contrast, $\chi_a$ remains the same for the defect atoms, indicating the $\chi_a$ of defect atoms to rapidly converge with regard to system size. This means that even using small defect cells, $\chi_a$ is a usefull measure to effectively determine the atoms belonging to the defect, and even to which degree.
This allows for the efficient calculation of the fragment spectra of a defect in a larger supercell system. Furthermore, the $\chi_a$ values also allow for the systematic improvement of such fragment spectra (\textit{cf.} below and Figure~\ref{fig_fragment}). Instead of defining a purely spatial threshold function, the threshold for inclusion can now be directly related to the atom's contribution to the defect spectrum.
Alternately, it is also possible to directly construct a defect spectrum from the atom projected spectra, using $\chi_a$ for selection purposes. The resulting spectrum will contain all (defect) features missing in the host spectrum. In addition, this defect spectrum will also contain contributions due to the interaction of the defect and the host lattice. These are incorporated through the (defect) atom projection of (delocalized) host system modes. This partitioning of the system into a host and defect fraction could be used as a platform to calculate derived thermodynamic contributions due to the defect (which goes beyond the scope of the current work).
Furthermore, because of the small size of the fragment to consider, one could more easily move beyond the standard harmonic approximation.\cite{PawelPT:JPhysChemLett2016,PicciniGM:JChemTheoryComp2014,DeWispelaereKEntropyTSmodel2018} \\
\section{Diamond based defects}
To evaluate our method, four different defects in diamond are considered.
\begin{itemize}
  \item The substitutional Eu dopant in diamond (Eu$_{sub}$): This heavy lanthanide dopant gives rise to low lying atomic phonon bands with a clearly distinguishable peak in the phonon spectrum.\cite{VanpouckeDannyEP:2019c_DRM_EuDope}
  \item The $\langle001\rangle$ split interstitial (C$_{i}$): This intrinsic defect places two C atoms at a single site. It provides a local breaking of the symmetry with limited change in the chemical environment. As a result, two very distinct optical phonon peaks are created well above the bulk spectrum.
  \item The neutral C vacancy (C$_{V}$): This intrinsic defect is obtained by removing a single C atom, and as a result, it resembles pristine diamond most closely. (The complex electronic structure and its influence on the defect geometry was modeled using the DFT+U method,\cite{VanpouckeDannyEP:2017d_DRM_DFTUVac} in contrast to the other defects where no +U correction on C was used.)
  \item The negatively charged nitrogen-vacancy centre (NV$^-$): One of the most discussed and studied defects in diamond. This defect presents a combination of a substitutional dopant and a carbon vacancy. Due to a mass comparable to that of C, the N atom gives rise to a spectrum comparable to that of C itself, making it challenging to extract while being of great interest for applications.
\end{itemize}

\subsection{Defect phonon spectra}
\begin{figure}[tb!]
\includegraphics*[width=\linewidth,keepaspectratio=true]{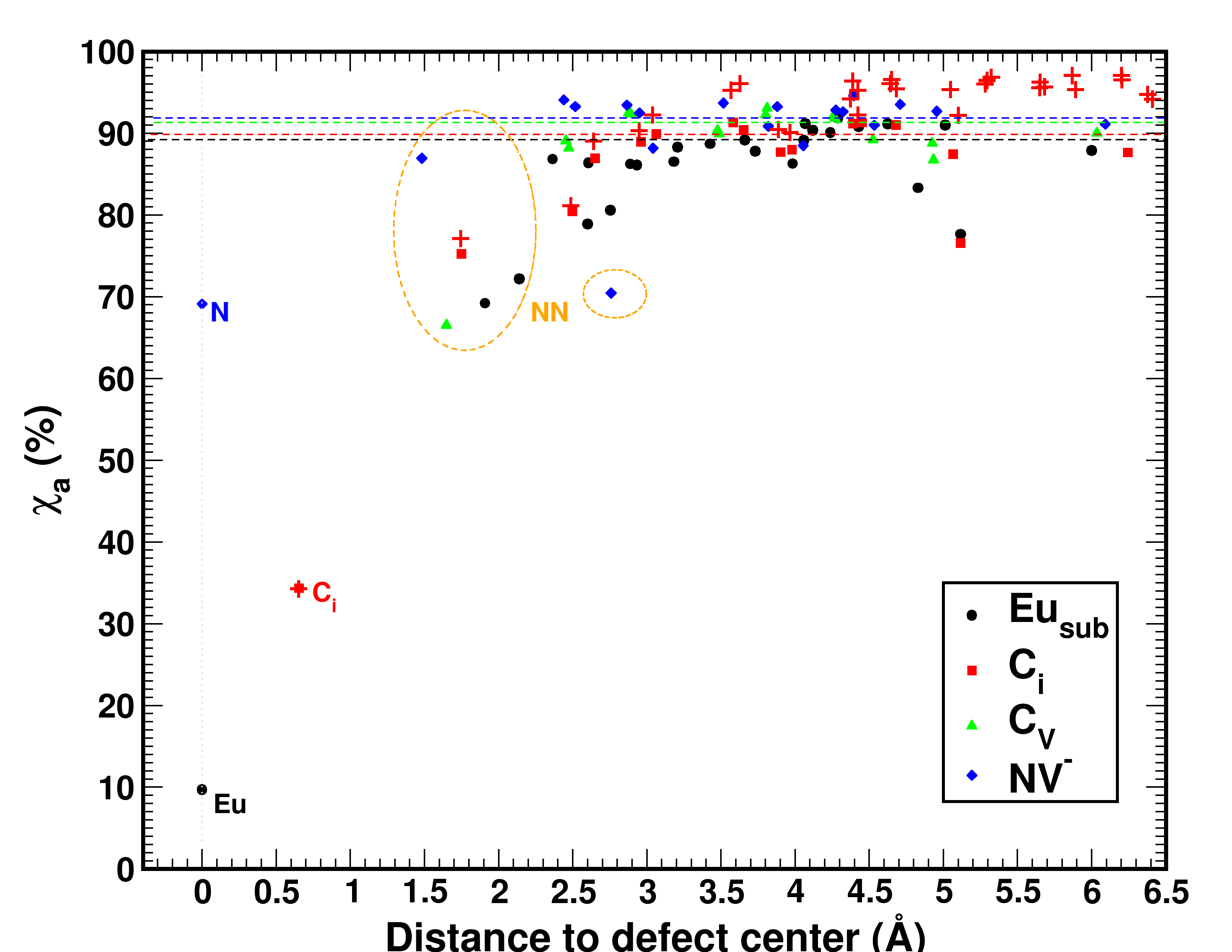}
\caption{%
    The overlap $\chi_a$ for each atom of the diamond defect systems modeled using a small $2\times 2\times 2$ conventional cell ($64$ atoms). The atoms forming the defect centers are indicated, as well as their nearest neighbours (NN). The horizontal dashed lines show the average $\chi_a$ value found for atoms in a range of $3$--$5$ \AA\ from the defect center for the $2\times 2\times 2$ defect cells. The red crosses show the result for a C$_i$ defect using a $3\times 3\times 3$ conventional cell (216 atoms), for comparison.\label{fig_overlapVSdist}
  }
\end{figure}
\begin{figure}[tb!]%
\includegraphics*[width=\linewidth,keepaspectratio=true]{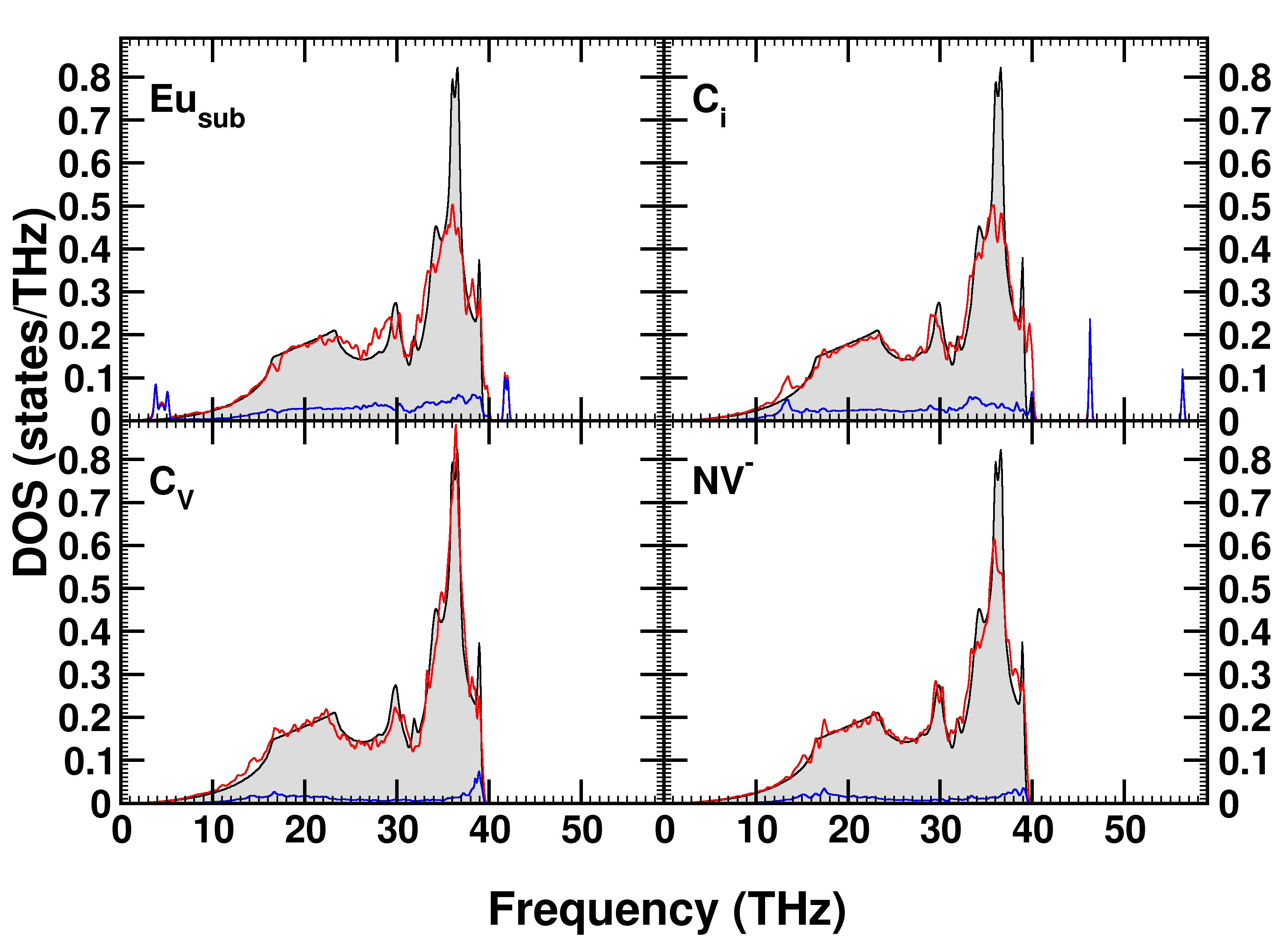}
\caption{%
  The phonon spectra of the four diamond defects systems (red curves) and the partial phonon spectrum due to the defect (blue curve) with a threshold of $\chi_a < 85$\%. The bulk diamond spectrum (black curve) is given as reference.\label{fig_spectra}
  }
\end{figure}

The overlap $\chi_a$ is calculated for each atom in the defect systems. In Figure~\ref{fig_overlapVSdist}, $\chi_a$ is shown as function of the distance of atom $a$ to the defect center. The atoms at the center of the defect are indicated, as well as the nearest neighbour (NN) atoms to the defect. In Figure~\ref{fig_spectra}, the phonon spectrum of the different defects is presented in comparison to the phonon spectrum of pristine diamond. Both the Eu$_{sub}$ and the C$_i$ defect give rise to clearly distinguishable phonon peaks, which show little to no overlap with the host phonon spectrum. This results in very low $\chi_a$ values for Eu and the interstitial C atoms, as is seen in Figure~\ref{fig_overlapVSdist}. The shells of NN and next-NN atoms already show rather large $\chi_a$ values of $70$--$80$\%, indicating that their atom-projected spectra are harder to distinguish from that of a host atom, but still clearly different. In contrast to these outspoken differences, the C$_V$ defect system shows a phonon spectrum quite similar to that of the host system, making it hard to indicate the differences and their sources. However, looking at Figure~\ref{fig_overlapVSdist}, four atoms stand out clearly with a $\chi_a \approx 70$\%: the four atoms surrounding the vacancy. Next-NN C atoms present a converged host character, showing this defect to be strongly localized on the vacancy and its four surrounding atoms. Considering the projected phonon DOS associated with these atoms (blue curve in the bottom left panel of Figure~\ref{fig_spectra}), it becomes clear that the defect spectrum consists of a peak at the high end of the spectrum, and two peaks around $14.5$ and $17$ THz. Turning our attention to the NV$^-$ defect, we notice in Figure~\ref{fig_overlapVSdist} that for the N atom, as well as for the three C atoms surrounding the vacancy, $\chi_a \approx 70$\%, similar as for the C$_V$ defect. The NN C atoms surrounding the N atom, on the other hand, present a high $\chi_a$ associated with host atoms. The defect spectrum is also quite similar to that of the C$_V$ defect, with additional peaks in the range of $14$--$17$ THz. However, in contrast to the C$_V$ defect, the peak at the high end of the spectrum is much less pronounced.\\
\subsection{Comparison to the fragment spectrum: the $\langle001\rangle$ split interstitial}
\begin{figure}[tb!]%
\includegraphics*[width=\linewidth,keepaspectratio=true]{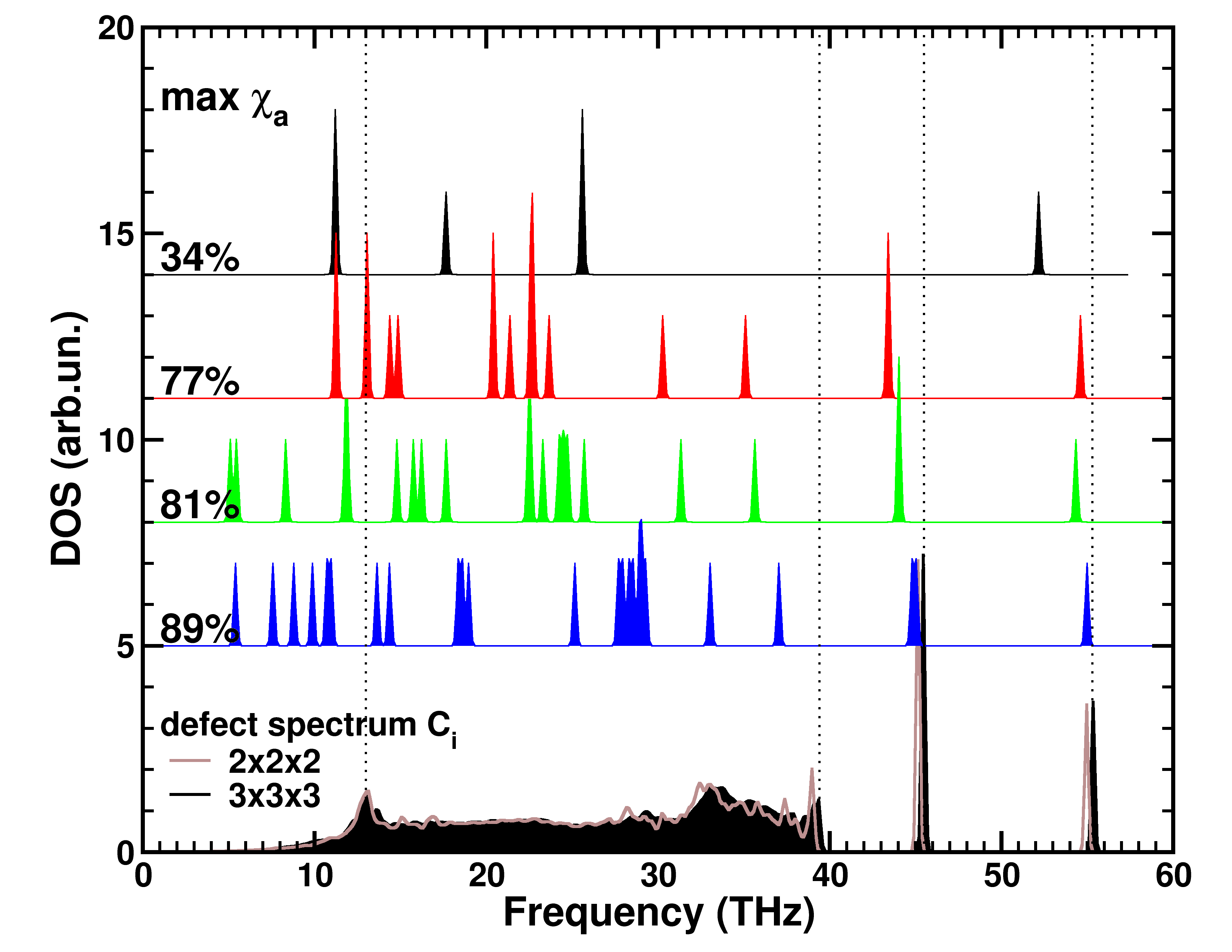}
\caption{%
  The phonon spectrum of the C$_{i}$ defect as obtained using a $65$ ($2\times 2\times 2$) and a 217 ($3\times 3\times 3$) atom supercell, with $\chi_a < 85$\%. Fragment spectra obtained using only atoms with $\chi_a < 34, 77, 81,$ and $89$\% in the $3\times 3\times 3$ system, are shown in comparison. Dotted lines indicate the position of specific defect spectrum features.
  \label{fig_fragment} }
\end{figure}
In Figure~\ref{fig_fragment} the defect spectrum of the C$_{i}$ defect is shown, obtained in both a \emph{smaller} $2\times 2\times 2$ (brown curve) and \emph{larger} $3\times 3\times 3$ (black) diamond supercell. It shows that the $2\times 2\times 2$ supercell is sufficient to construct a well-converged defect spectrum. The two optical peaks at about $45$ THz and $55$ THz are found to be within $0.5$ THz of the results obtained with the $3\times 3\times 3$ supercell, while the feature at $13$ THz shows no visible deviation. Furthermore, the broad band, due to defect--host system interactions, is well converged. It is important to note that the computational cost for generating the first-principles Hessian matrix within a periodic plane waves approach (shown in Table~\ref{tab_CPU}), for the $2\times 2\times 2$ supercell is $24\times$ lower than for the larger supercell, making this a cost-efficient approach.\\
The defect spectrum is also compared to different fragment spectra obtained using the $3\times 3\times 3$ supercell. The atoms belonging to the fragment are determined using their $\chi_a$ value: $\chi_a \leq$ $34$\% ($2$ atoms), $77$\% ($6$ atoms), $81$\% ($10$ atoms), and $89$\% ($18$ atoms). The resulting defect spectra obtained using the fragment approach are shown in Figure~\ref{fig_fragment}. All fragments (except the smallest $2$-atom fragment) give rise to the two optical modes, and it is only the largest fragment which positions them with an accuracy comparable to the $2\times 2\times 2$ defect spectrum (at almost twice the computational cost). More interestingly, the feature at $13$ THz is not retrieved in the fragment spectra, neither is the broad interaction band.
\begin{table}
  \caption{\label{tab_CPU} CPU time (days) required to generate the Hessian matrix of the C$_{i}$ defect in diamond, using first-principles quantum mechanical calculations.}
  \begin{ruledtabular}
  \begin{tabular}{@{}l|r@{}}
    \hline
    system & CPU time (days) \\
    \hline
    $3\times 3\times 3$ full spectrum  & 3322   \\
    $2\times 2\times 2$ full spectrum  &  137   \\
    fragment 34\% & 32 \\
    fragment 77\% & 83 \\
    fragment 81\% & 138 \\
    fragment 89\% & 251 \\
    \hline
  \end{tabular}
  \end{ruledtabular}
\end{table}
\section{Conclusions}
In this work, a method is presented for determining the phonon-spectrum of a \emph{defect} using relatively small periodic first-principles calculations. Our method provides a quantitative measure for assigning atoms to a defect. This allows it to be used in tandem with a fragment approach to efficiently obtain incrementally more accurate fragments in much larger supercells.
Alternately, combining the atom projected vibrational spectra of the \textit{defect} atoms gives rise to a quickly converging defect spectrum which combines the defect specific features of the spectrum with the contributions due to defect-host interactions.
The resulting partitioning of the system spectrum into a host and defect component opens up the possibility for similar partitioning of properties derived from the phonon spectrum, which is the subject of ongoing research. The presented methods have been implemented in the HIVE package.\cite{HIVE4}

\acknowledgements{
The computational resources and services used in this work were provided by the VSC (Flemish Supercomputer Center), funded by the Research Foundation Flanders (FWO) and the Flemish Government--department EWI. I thank A. Vanwetswinkel for his contribution to the implementation of the atom projection scheme.
}


\begin{thebibliography}{35}
\bibitem[{not()}]{note_EigVectNorm}
\bibinfo{note}{In numerical packages, the obtained eigenvectors, from for
  example a singular value decomposition, will be normalized by default.}
\end{thebibliography}

\end{document}